\documentclass[aps,prl,twocolumn,superscriptaddress,nofootinbib]{revtex4-2}
\usepackage{amsmath,amssymb,graphicx}
\usepackage[colorlinks=true,linkcolor=blue,urlcolor=blue,citecolor=blue]{hyperref}
\usepackage{orcidlink}

\begin{document}
\title{A Dark-Sector Mass Flip-Flop Mechanism for the ANITA-IV Events}

\author{Dibya~S.~Chattopadhyay~\orcidlink{0000-0003-2323-3950}}
\email{dibya.chattopadhyay@okstate.edu}
\affiliation{Department of Physics, Oklahoma State University, Stillwater, OK 74078, USA}

\author{Carlos~A.~Arg\"uelles~\orcidlink{0000-0003-4186-4182}}
\email{carguelles@fas.harvard.edu}
\affiliation{Department of Physics \& Laboratory for Particle Physics and Cosmology, Harvard University, Cambridge, MA 02138, USA}

\author{Vedran Brdar~\orcidlink{0000-0001-7027-5104}}
\email{vedran.brdar@okstate.edu}
\affiliation{Department of Physics, Oklahoma State University, Stillwater, OK 74078, USA}

\begin{abstract}
ANITA-IV has observed four upward-going air-shower candidates just below the horizon, consistent with ultra-high-energy $\nu_\tau$, yet IceCube, despite its much larger exposure, has detected no neutrino events at such energies.
We propose a dark-sector mass flip-flop mechanism that reconciles these observations.
We consider an all-sky, monoenergetic flux of ultra-high-energy $N_1$, that is lighter than its partner $N_2$ and stable in vacuum.
In our scenario, an ultralight scalar mediating interactions with Earth matter inverts the mass ordering of $N_1$ and $N_2$, allowing an incident $N_1$ to decay into $N_2$ deep underground.
The resulting $N_2$ propagates unimpeded through matter and exits the Earth, where the decay $N_2 \to N_1 Z'$ becomes kinematically allowed in the atmosphere, initiating an extensive air shower.
At IceCube depths, neither $N_1\to N_2 Z'$ nor $N_2\to N_1 Z'$ is kinematically allowed, which explains the absence of analogous events at IceCube.
Normalizing the $N_1$ flux to the four ANITA-IV events, we also predict three events during PUEO's first flight.
\end{abstract}
\maketitle

\textbf{Introduction.} 
During the fourth flight of the Antarctic Impulsive Transient Antenna (ANITA-IV)~\cite{ANITA:2008mzi,ANITA:2019wyx}, four cosmic-ray-like events, with non-inverted polarity, were observed just below the horizon~\cite{ANITA:2020gmv}. Their properties are compatible with upgoing extensive air showers (EAS), with a possible origin being $\tau$ leptons produced by ultra-high-energy (UHE) $\nu_\tau$ traversing Earth~\cite{ANITA:2021xxh}. This interpretation, however, faces a severe challenge from IceCube's~\cite{IceCube:2006tjp,IceCube:2016zyt} non-observation of a corresponding event population~\cite{IceCubeCollaborationSS:2025jbi,Bertolez-Martinez:2023scp,Arguelles:2025ewg}. A joint analysis of ANITA-IV and IceCube quantifies the severity of this discrepancy: the best-fit diffuse neutrino flux predicts approximately four events at IceCube while resulting in an expected event count of $\mathcal{O}(10^{-3})$ at ANITA-IV. This corresponds to a tension exceeding $7\sigma$~\cite{Chattopadhyay:2026yoa}. Considering rare transient sources partially alleviates the discrepancy; however, the tension remains at more than $5\sigma$ in such scenarios~\cite{Chattopadhyay:2026yoa}. This tension motivates beyond-the-Standard-Model (BSM) explanations that can account for the ANITA-IV air showers while evading the constraints from IceCube~\cite{Bertolez-Martinez:2023scp,Arguelles:2025ewg} and  the Pierre Auger Observatory.

\begin{figure}[t!]
\centering
\includegraphics[width=\linewidth]{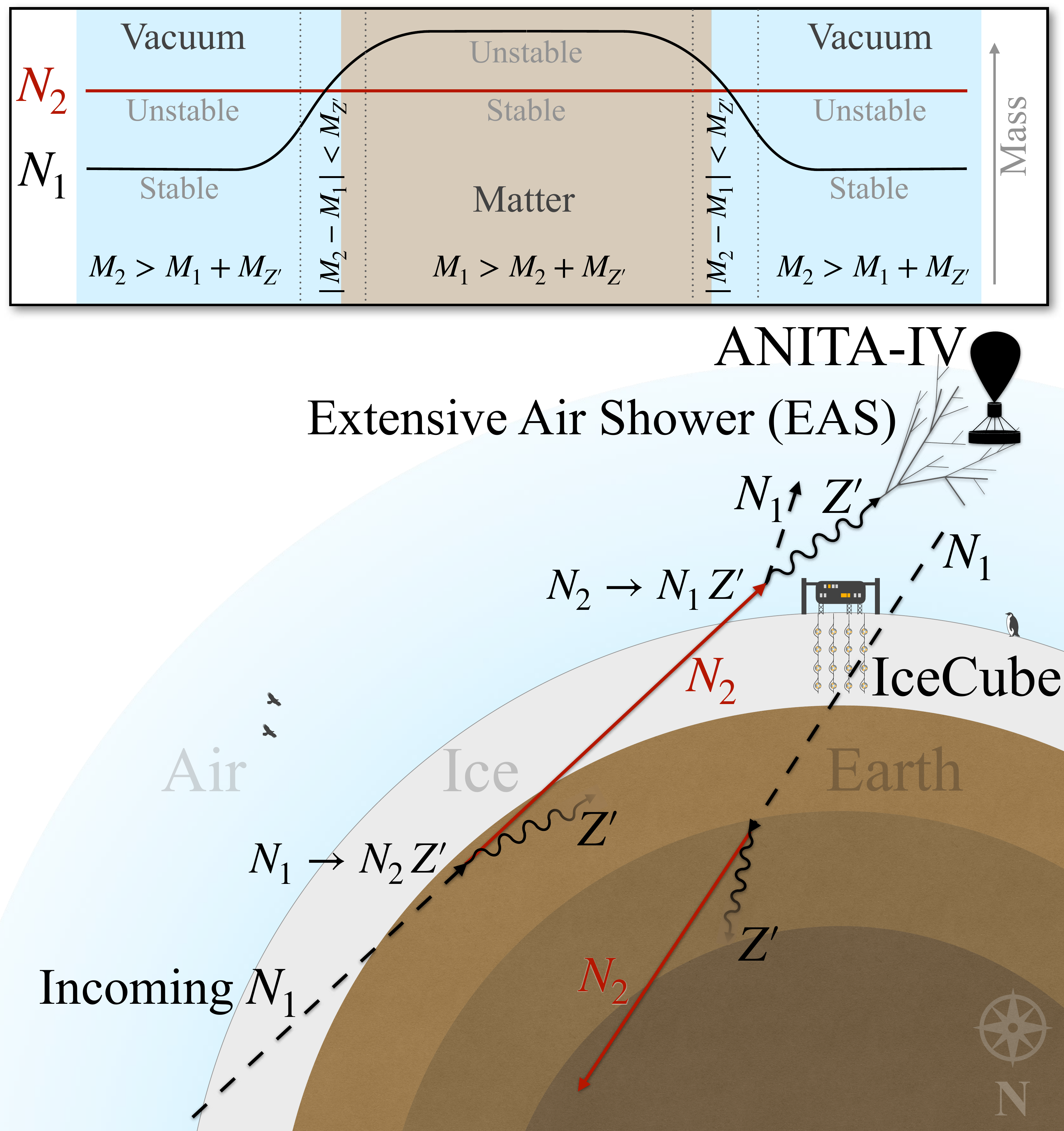}
\caption{The dark-sector mass flip-flop mechanism. The upper panel shows the reversal of the $N_1$ and $N_2$ mass ordering and the regions where $|M_2-M_1|\leq M_{Z'}$ blocks both $N_1$ and $N_2$ decays. The lower panel shows an upgoing $N_2$ trajectory that produces an atmospheric shower and a downgoing $N_1$ trajectory that crosses IceCube before decaying to $N_2$.}
\vspace{-0.4cm}
\label{fig:schematic}
\end{figure}

In this Letter, we propose a matter-induced reversal of a dark-fermion mass ordering, which we refer to as the mass flip-flop.
We consider two dark fermions, $N_1$ and $N_2$, with masses $M_1 < M_2$ in vacuum, so that $N_1$ is stable.
Coherent forward scattering of $N_1$ off nucleons in Earth mediated by an ultralight scalar, $\phi$, adds a matter-dependent contribution to its effective mass $M_1$~\cite{Davoudiasl:2023uiq,Davoudiasl:2026mqk}, which exceeds $M_2$ deep inside Earth.
We further assume an incoming flux of UHE $N_1$.
A possible source of such $N_1$ is the two-body decay of EeV-scale dark matter~\cite{Chung:1998zb,Chung:1998ua,Heurtier:2019git,Hooper:2019ytr,Heurtier:2019rkz,Morrison:2022zwk,Kolb:2023ydq,Narita:2025udw}, which yields a nearly monoenergetic spectrum.
An incident $N_1$, stable in vacuum, can thus decay via $N_1\to N_2 \, Z'$, where $Z'$ is a GeV-scale vector boson, deep underground~\cite{Davoudiasl:2026mqk}. The resulting $N_2$ then emerges from Earth and, at sufficient altitude, undergoes the reverse decay $N_2\to N_1\,Z'$. The produced $Z'$ subsequently decays hadronically~\cite{Pospelov:2011ha,Tulin:2014tya}, initiating an air shower.
Crucially, the $\mathcal{O}({\rm GeV})$ $Z'$ mass, $M_{Z'}$, separates the two decay thresholds, leaving an intermediate layer in which $|M_2-M_1|\leq M_{Z'}$ blocks both $N_1$ and $N_2$ decay channels.
If this layer encompasses IceCube, the same incident flux can produce air showers visible to ANITA-IV while yielding no direct signal in IceCube.
In Figure~\ref{fig:schematic}, we illustrate this mechanism for both upgoing and downgoing trajectories.

In what follows, we present the model and the respective $N_1$ and $N_2$ matter-dependent masses, which allow us to derive the conditions for decay blocking in the vicinity of IceCube. We then calculate the ANITA-IV effective area and normalize the flux to reproduce the four ANITA-IV candidate events. Finally, we extend this analysis to the Payload for Ultrahigh Energy Observations (PUEO) and find that its first flight should detect $\sim 3$ events.

\textbf{Mass flip-flop and decay blocking.} We introduce the Lagrangian~\cite{Pospelov:2011ha,Tulin:2014tya,Davoudiasl:2023uiq,Davoudiasl:2026mqk}
\begin{align}
\label{eq:lagrangian}
\mathcal L& \supset \!\sum_{i=1}^2\overline N_i \!\left(i \!\not\!\partial-M_i^{(0)}\right)\! N_i +\tfrac12(\partial\phi)^2-\tfrac12m_\phi^2\phi^2\\
&\; \; -\phi\bigg(y_1\overline N_1N_1
+\!\sum_{f\in{\rm mat}}\! y_f\overline f f\bigg)-\tfrac14 Z'_{\mu\nu}Z'^{\mu\nu}\nonumber\\
&\; \; +\tfrac12M_{Z'}^2Z'_\mu Z'^\mu \! + g_BZ'_\mu J_B^\mu+\!\Big(g_{12}Z'_\mu\overline N_2\gamma^\mu N_1 \! +\mathrm{h.c.}\!\Big), \nonumber
\end{align}
where $M_1^{(0)}$ and $M_2^{(0)}$ are the dark fermion masses in vacuum, the sum $\sum_{f \in {\rm mat}}$ runs over the fermions in ordinary matter, $Z'_{\mu\nu} \equiv \partial_\mu Z'_\nu-\partial_\nu Z'_\mu$, and
$J_B^\mu=\frac13\sum_q\overline q\gamma^\mu q$ is the baryon current. The $U(1)_B$ mediator $Z'$ therefore couples to quarks with strength $g_B/3$.
We assume that $\phi$ couples to $N_1$ and does not interact with $N_2$. The mass hierarchy in vacuum, $M_2^{(0)}>M_1^{(0)}+M_{Z'}$, allows $N_2\to N_1\,Z'$ while leaving $N_1$ stable.

Earth matter induces a position-dependent mass correction for $N_1$~\cite{Smirnov:2019cae,Babu:2019iml,Davoudiasl:2023uiq}:
\begin{align}
M_1(\mathbf x)&=M_1^{(0)}+\delta M_1(\mathbf x),
\qquad M_2=M_2^{(0)},\nonumber\\
\delta M_1(\mathbf x)&=-y_1\sum_f y_f
\int d^3x'\,
\frac{n_f(\mathbf x')e^{-m_\phi|\mathbf x-\mathbf x'|}}
{4\pi|\mathbf x-\mathbf x'|},
\label{eq:mass}
\end{align}
where $n_f$ is the number density of fermion $f$, and we use $n_f\simeq\langle\overline f f\rangle$, valid for nonrelativistic matter.
We choose $y_1\,y_N<0$, where $y_N$ is the coupling of $\phi$ to nucleons. With this choice, Earth matter raises $M_1$ while leaving $M_2$ unchanged.
A sufficiently large mass correction reverses the vacuum mass ordering and opens $N_1\to N_2\, Z'$ deep underground.

\begin{figure}[t!]
\centering
\includegraphics[width=0.9\linewidth]{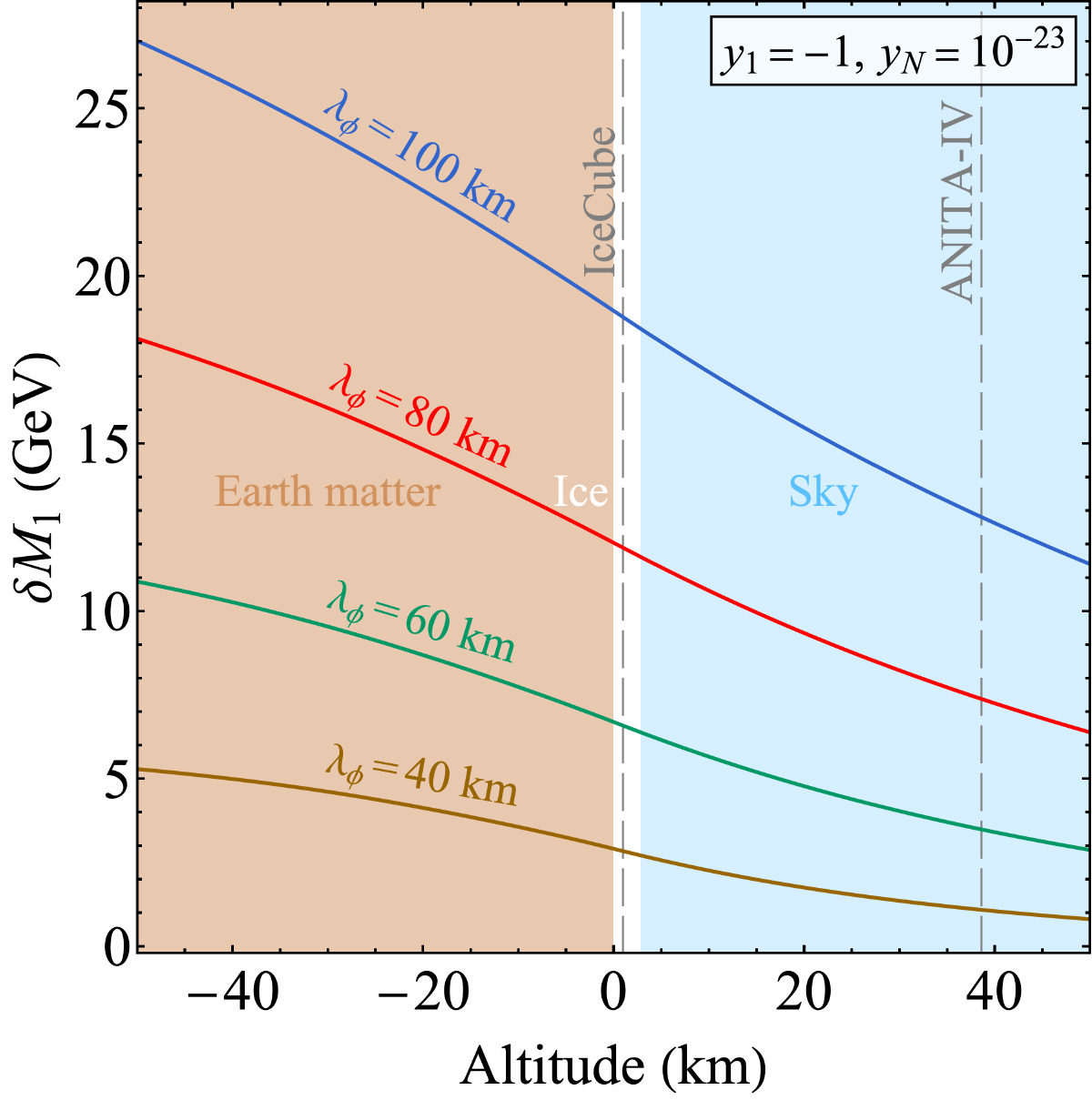}
\vspace{-0.1cm}
\caption{Scalar-induced mass correction $\delta M_1$ as a function of altitude for various interaction ranges, with $y_1=-1$ and $y_N=10^{-23}$. The brown, white, and blue regions denote Earth matter, ice, and atmosphere, respectively, while dashed vertical lines mark the altitudes of IceCube and ANITA-IV.}
\vspace{-0.2cm}
\label{fig:microphysics}
\end{figure}

The variation of the mass correction, $\delta M_1$, with depth and altitude determines where the two decay channels are allowed.
We evaluate Eq.~\eqref{eq:mass} using the matter density distribution in Earth and ice, accounting for the finite scalar interaction range $\lambda_\phi=m_\phi^{-1}$. Outside Earth, the correction falls as $\sim \exp(-r/\lambda_\phi)/r$, allowing the \mbox{$N_2 \to N_1 \, Z'$} decay to open at a finite altitude. Figure~\ref{fig:microphysics} shows $\delta M_1$ across Earth matter, ice, and atmosphere for $y_1=-1$, $y_N=10^{-23}$, and for several values of $\lambda_\phi$. Longer interaction ranges produce larger corrections farther into the atmosphere. 

The decay $N_1\to N_2\,Z'$ requires $M_1>M_2+M_{Z'}$, while the reverse decay requires $M_1<M_2-M_{Z'}$. Both channels are therefore kinematically blocked when
\begin{equation}
|M_1(h)-M_2|\leq M_{Z'},
\label{eq:block}
\end{equation}
where $h$ denotes altitude above sea level. Therefore, the nonzero $Z'$ mass separates the two kinematic regions corresponding to underground conversion of $N_1 \to N_2$, and atmospheric decay of $N_2 \to N_1 \, Z'$.

\begin{figure}[t!]
\centering
\includegraphics[width=0.9\linewidth]
{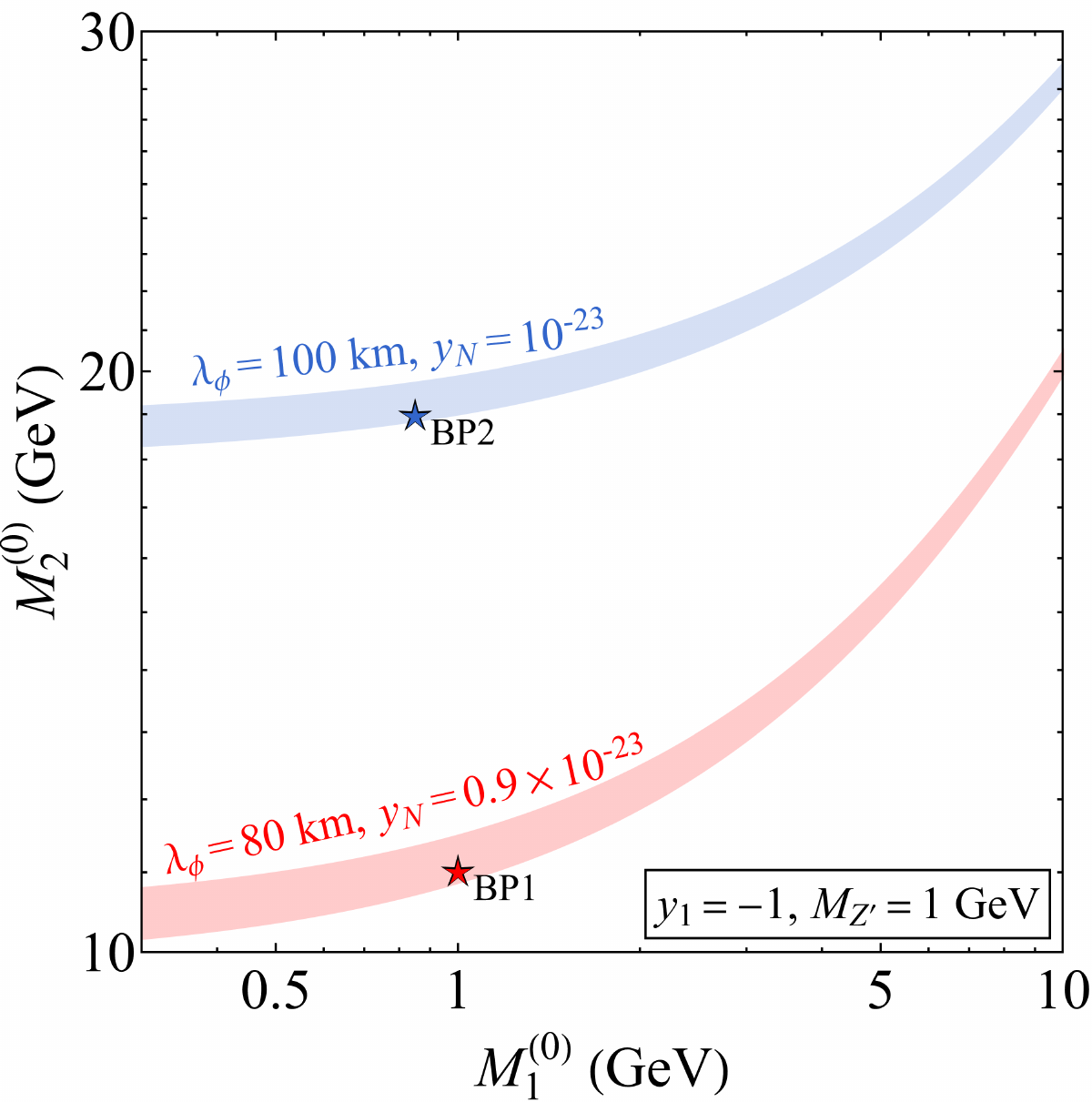}
\caption{Regions in the $(M_1^{(0)}\!,\,M_2^{(0)})$ parameter space for which the lower and upper boundaries of the decay-blocked layer satisfy $-5\,{\rm km}<h_-<0\,{\rm km}$ and $2\,{\rm km}<h_+<20\,{\rm km}$, respectively, encompassing all of IceCube. The red and blue stars mark the benchmark points BP1 and BP2 (see Table~\ref{tab:benchmarks}). We fix $M_{Z'}=1\,{\rm GeV}$ and $y_1=-1$.}
\label{fig:microphysics2}
\end{figure}

We denote the lower and upper boundaries of the intermediate `decay-blocked' region by $h_-$ and $h_+$, respectively, defined by $M_1(h_\pm)=M_2\mp M_{Z'}$. Below $h_-$, an incident $N_1$ can decay into $N_2$, so only trajectories that reach this depth can contribute to the $N_1 \to N_2 \to {\rm EAS}$ mechanism. Above $h_+$, the reverse decay becomes possible. A higher $h_+$ therefore leaves less atmospheric column density for the air shower to develop.
In Figure~\ref{fig:microphysics2}, we show the vacuum masses of $N_1$ and $N_2$ for which $-5\,{\rm km}<h_-<0\,{\rm km}$ and $2\,{\rm km}<h_+<20\,{\rm km}$. Here, the altitude $h$ is measured from sea level, so IceCube's instrumented volume, located $1.45$--$2.45\,{\rm km}$ below the ice surface at the South Pole, spans $h\simeq0.4$--$1.4\,{\rm km}$~\cite{IceCube:2016zyt}.
These windows place IceCube inside the blocked region and restrict the altitude at which the reverse decay opens. For a given $\lambda_\phi$ and $y_N$, the threshold locations depend on the vacuum masses only through the splitting $M_2^{(0)}\! -\! M_1^{(0)}$, giving rise to the bands in the figure. The stars denote the two benchmark points listed in Table~\ref{tab:benchmarks}.

For both benchmark points, underground conversion opens at nearly the same altitude, $h_-\simeq-1\,{\rm km}$, and both $N_1$ and $N_2$ decays remain blocked throughout IceCube's instrumented volume. The reverse decay opens at about $15\,{\rm km}$ for BP1 and $10\,{\rm km}$ for BP2. An emerging $N_2$ can therefore initiate a shower lower in the atmosphere for BP2, where more air remains along its path. This difference is especially relevant for steeper trajectories, which traverse the atmosphere over a shorter distance.

\begin{table}[t!]
\centering
\renewcommand{\arraystretch}{1.12}
\setlength{\tabcolsep}{4pt}
\begin{tabular}{|l|c|c|c|c|c|c|}
\hline
 & $M_1^{(0)}$ & $M_2^{(0)}$ & $\lambda_\phi$ & $y_N$
 & $h_-$ & $h_+$ \\
 & (GeV) & (GeV) & (km) & $(10^{-23})$ & (km) & (km) \\
\hline
BP1 & 1.00 & 11.0 & 80  & 0.9 & $-1.15$ & 14.77 \\
\hline
BP2 & 0.85 & 19.0 & 100 & 1.0 & $-0.96$ & 9.95  \\
\hline
\end{tabular}
\caption{Considered benchmark points and the resulting decay-blocked region, denoted by $h_-$ and $h_+$. We fix $M_{Z'}=1\,{\rm GeV}$ and $y_1=-1$ for both benchmark points.}
\label{tab:benchmarks}
\vspace{-0.3cm}
\end{table}

For the chosen benchmark couplings (see Table~\ref{tab:benchmarks}), two-body decays can occur over terrestrial distances, but both channels are blocked throughout IceCube. The effects of three-body decays and vector-mediated scattering are negligible, leaving no appreciable cascade signal inside the IceCube detector. In principle, a decay outside the IceCube array could still produce a through-going $\mu$ or $\tau$-lepton track~\cite{Dutta:2000hh,GarciaCanal:2012qth,Gaisser:2016uoy}. However, atmospheric $N_2\to N_1\,Z'$ showers point away from Earth and therefore happen after $N_2$ crosses IceCube, whereas downgoing $N_1$ can convert only below the decay-blocked region, i.e., below IceCube. The relevant tracks can therefore arise only from upgoing $N_1$ trajectories that pass below $h_-$. Along such trajectories, $N_1\to N_2\,Z'$ typically occurs while $N_1$ is still far from IceCube. For both BP1 and BP2, with $g_{12}=10^{-5}$, the $N_1$ flux required to explain four events at ANITA-IV gives an expectation value well below 1 event for $Z'$ decay-induced showers originating within $100\,{\rm km}$ from IceCube over the last 15 years. The same benchmarks also suppress direct decays at KM3NeT/ARCA~\cite{KM3Net:2016zxf}. Approximating its surroundings with a $3.45\,{\rm km}$ ocean layer, we obtain $h_-\simeq-3.6\,{\rm km}$ and $-3.4\,{\rm km}$ (with $h_+ \simeq 12.3\,{\rm km}$ and $7.5\,{\rm km}$) for BP1 and BP2, respectively. The blocked layers encompass ARCA's instrumented depths, though the BP2 boundary lies near the array's lower edge. A more precise treatment requires accounting for the local bathymetry.

Similar arguments apply to the Pierre Auger Observatory.
Since $N_2$ is unstable in vacuum, the incident flux consists solely of $N_1$, which cannot decay above $h_-$; downgoing showers are therefore absent, leaving cosmic-ray observables and searches for downgoing neutrinos~\cite{PierreAuger:2019ens} unaffected.
Upgoing $N_2\to N_1\,Z'$ decays occur above $h_+$ and produce showers that point away from Earth and therefore cannot deposit particles in the surface array, located at $\simeq 1.4\,{\rm km}$ above sea level, making the Earth-skimming $\nu_\tau$ search~\cite{PierreAuger:2019ens} insensitive.
The fluorescence detector, however, observes isotropically emitted fluorescence light and therefore remains sensitive to such showers regardless of their direction.
Its dedicated search for upgoing showers arriving from more than $20^\circ$ below the horizon found one candidate with an expected background of $0.27\pm0.12$~\cite{PierreAuger:2025hvl}.
The exposure of this search, however, falls rapidly with the shower starting altitude.\footnote{The Pierre Auger simulations cover decay altitudes of up to $9\,{\rm km}$ above ground, i.e., $10.4\,{\rm km}$ above sea level~\cite{PierreAuger:2025hvl}.}
Since $N_2\to N_1\,Z'$ decays start only at higher altitudes for both BP1 and BP2, we expect a substantially reduced event count at the Pierre Auger Observatory; a quantitative prediction would require knowledge of the fluorescence detector's response to showers starting at higher altitudes.

The position-dependent decay rates and energy distributions along the $N_1\to N_2\to{\rm EAS}$ chain determine where atmospheric showers form and how much energy they carry. Our semi-analytic ANITA-IV effective area calculations incorporate the underground conversion and atmospheric decay probabilities, and adapt the detector response derived for Standard Model $\nu_\tau$-induced showers by matching the visible shower energy and correcting for the different branching fractions and shower-development criteria. The decay rates, energy transport, and effective-area calculation are detailed in Secs.~\ref{app:1} and~\ref{app:2} of the Supplemental Material.

\begin{figure*}[t!]
\centering
\includegraphics[width=0.486\linewidth]{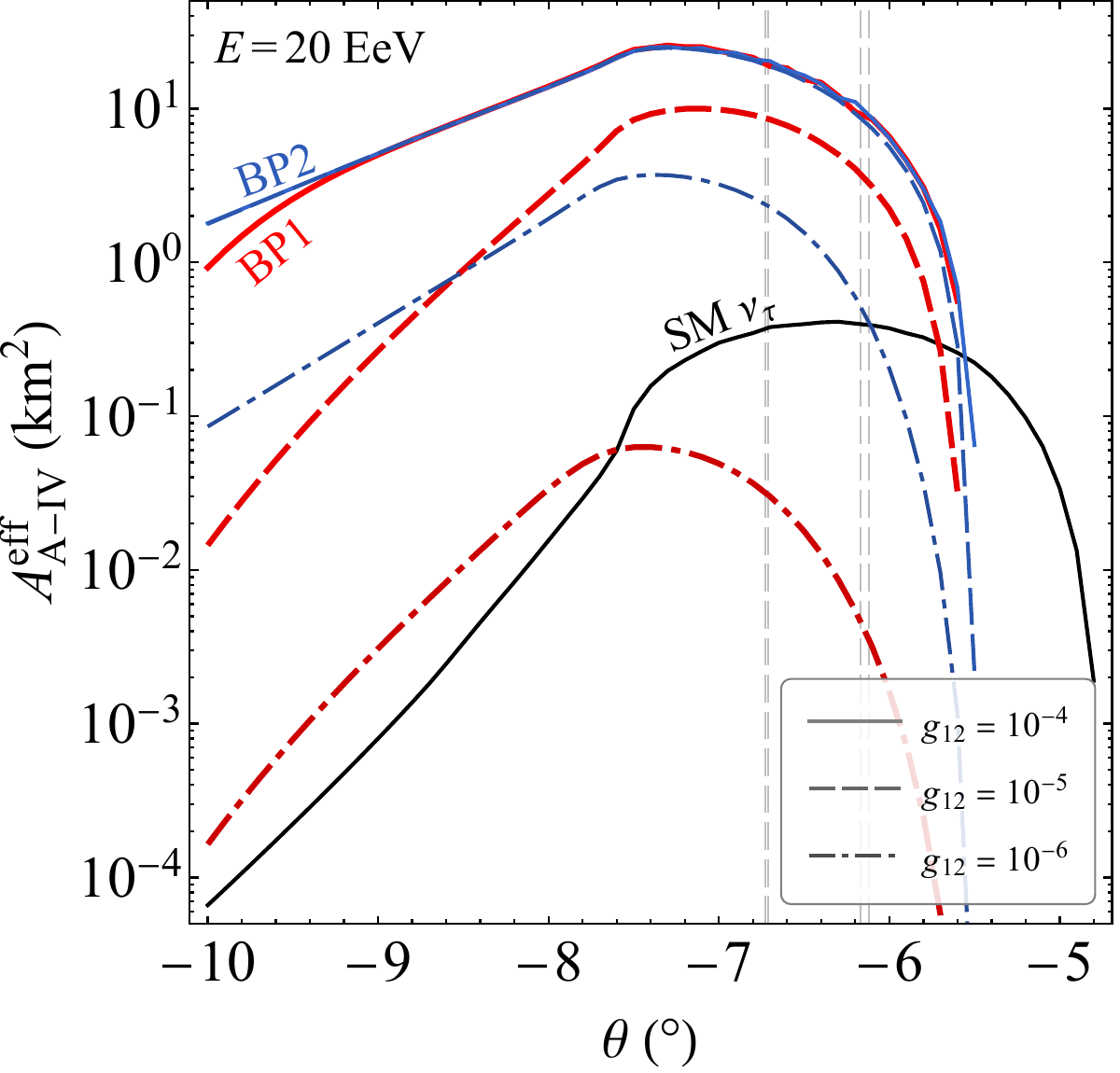}
\hfill
\includegraphics[width=0.465\linewidth]{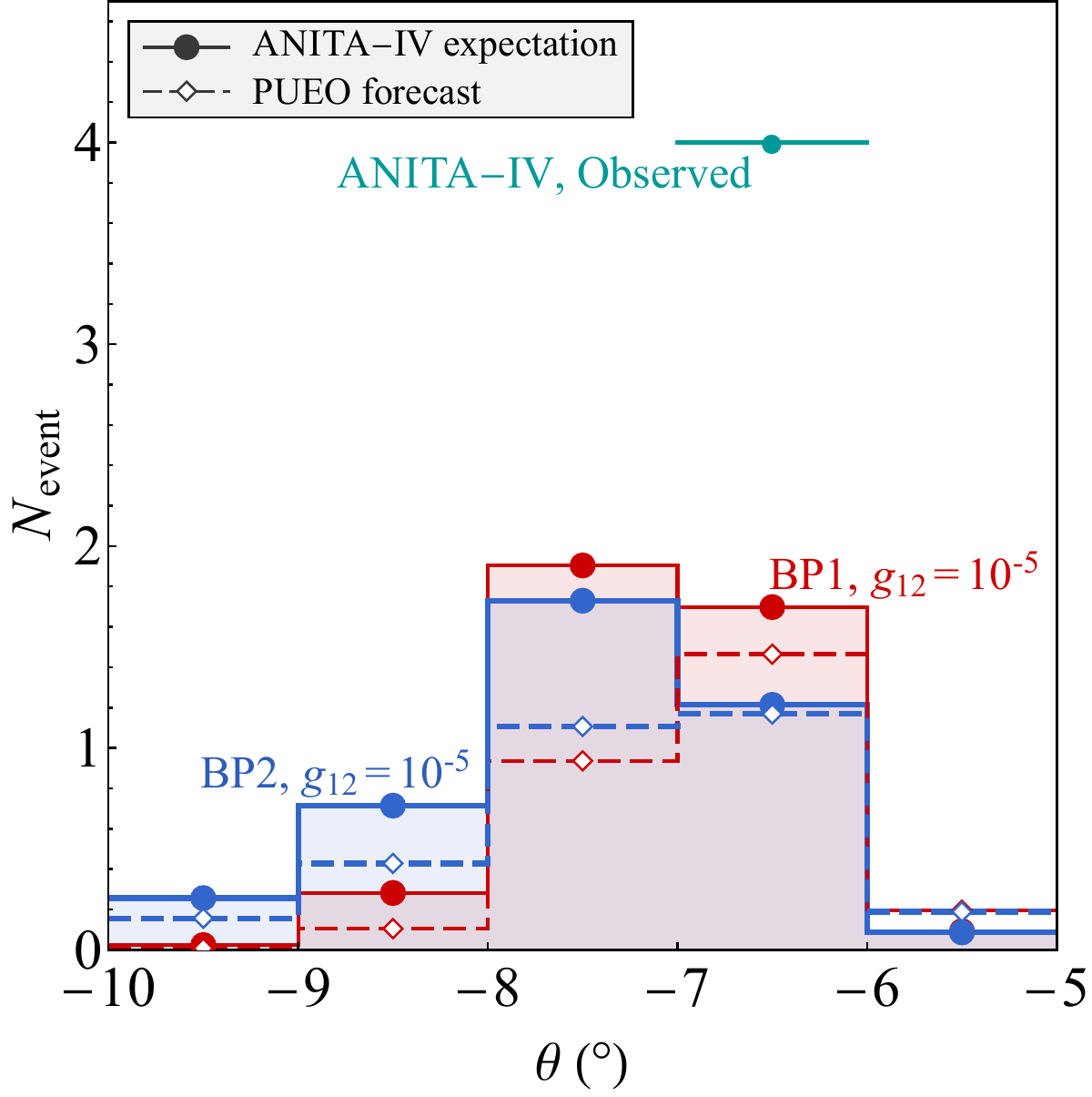}
\vspace{-0.1cm}
\caption{\textit{Left:} ANITA-IV effective area shown as a function of the elevation angle $\theta$ for an incident energy of $20\,{\rm EeV}$. The black curve shows the prediction for Earth-skimming $\nu_\tau$. Red and blue curves show the predictions for an incident $N_1$, for BP1 and BP2 (see Table~\ref{tab:benchmarks}), respectively, with solid, dashed, and dash-dotted lines corresponding to $g_{12}=10^{-4}$, $10^{-5}$, and $10^{-6}$. The four thin vertical dashed lines correspond to the event directions of the four anomalous ANITA-IV events. \textit{Right:} Expected event counts in $1^\circ$ bins, for ANITA-IV and PUEO, with $g_{12}=10^{-5}$. For each benchmark, we normalize the $N_1$ flux to a total of four ANITA-IV events (solid lines, with circles) and use the same flux to predict the yield of PUEO's first flight (denoted in dashed lines, with diamonds). The dark cyan point marks the four observed ANITA-IV events, all lying in the $(-7^\circ,-6^\circ)$ bin.}
\vspace{-0.2cm}
\label{fig:effarea}
\end{figure*}

\textbf{Results.} In the left panel of Figure~\ref{fig:effarea}, we compare the Standard Model (SM) effective area~\cite{Chattopadhyay:2026yoa} (black) with the corresponding effective areas for BP1 (red) and BP2 (blue); the solid, dashed, and dash-dotted curves correspond to $g_{12}=10^{-4}$, $10^{-5}$, and $10^{-6}$, respectively. The incoming $N_1$ energy is taken to be $E_1=20\,{\rm EeV}$.
The angular dependence is regulated by two competing requirements: shallow trajectories must reach sufficient depth for $N_1 \to N_2$, while steeper trajectories must provide enough atmospheric column for shower development above $h_+$.
The higher decay threshold in BP1, $h_+\simeq15\,{\rm km}$ compared with $10\,{\rm km}$ in BP2, therefore produces a faster decline toward steeper arrival directions. Increasing $g_{12}$ from $10^{-5}$ to $10^{-4}$ raises the BP1 peak from $10$ to $26\,{\rm km^2}$ but leaves the BP2 peak nearly unchanged.
Near the arrival direction of the four ANITA-IV events~\cite{ANITA:2020gmv}, both benchmarks yield effective areas that are 1--2 orders of magnitude above the SM prediction.

To estimate the event yield, we consider an isotropic flux of $N_1$ with a single incident energy, $d\Phi/dE=\Phi_0\delta(E-E_1)$, and take $E_1=20\,{\rm EeV}$ and $g_{12}=10^{-5}$ for both benchmarks. We require four ANITA-IV events~\cite{ANITA:2020gmv} within the near-horizon interval $\mathcal{H}=(-10^\circ,-5^\circ)$:
\begin{equation}
2\pi \, T_{\rm A-IV} \, \Phi_0
\int_{\mathcal{H}} A_{\rm eff}^{\rm A-IV}(E_1,\theta)\cos\theta\,d\theta = 4.
\label{eq:counts}
\end{equation}
Here, $\theta$ is the elevation angle relative to the horizon and $T_{\rm A-IV}$ is the ANITA-IV live time.
Integrating the effective area over $\mathcal{H}$ gives $1.7\,{\rm km^2\,sr}$ for BP1 and $5.4\,{\rm km^2\,sr}$ for BP2. For the $24.5$ live days of ANITA-IV~\cite{ANITA:2020gmv}, BP1 therefore requires a best-fit flux of $1.1\times10^{-16}\,{\rm cm^{-2}\,s^{-1}\,sr^{-1}}$, whereas the larger acceptance of BP2 gives the same yield for a roughly three times smaller flux of $3.5\times10^{-17}\,{\rm cm^{-2}\,s^{-1}\,sr^{-1}}$.

With the flux normalized to four ANITA-IV events, both benchmarks yield a near-horizon air-shower signal with both $N_1$ and $N_2$ decay channels blocked at IceCube. The mass flip-flop mechanism can thus accommodate the four ANITA-IV candidates~\cite{ANITA:2020gmv} together with the absence of a corresponding IceCube signal~\cite{IceCubeCollaborationSS:2025jbi,Bertolez-Martinez:2023scp,Chattopadhyay:2026yoa}, resolving the anomaly.
In the right panel of Figure~\ref{fig:effarea}, we show the expected number of events in each angular bin near the horizon. For BP1 and BP2, most of the expected near-horizon events fall between $-8^\circ$ and $-6^\circ$ in elevation. 

The mass flip-flop mechanism must also reproduce the observed angular distribution through the $N_1\to N_2\to{\rm EAS}$ process: ANITA-IV observed all four events in the $(-7^\circ,-6^\circ)$ bin~\cite{ANITA:2020gmv}.
To quantify the agreement, we compare the predicted and observed event numbers in five $1^\circ$ bins between $-10^\circ$ and $-5^\circ$, using an exact Poisson-deviance (likelihood-ratio) test~\cite{Baker:1983tu} conditional on four events, evaluated over all 70 possible arrangements of the events among the bins.
This gives $p=0.098$ for BP1 and $p=0.023$ for BP2, corresponding to one-sided Gaussian significances of $1.3\sigma$ and $2.0\sigma$, respectively~\cite{ParticleDataGroup:2024cfk}. BP1 is therefore consistent with the observed angular distribution, while BP2 is in mild tension with it. 
A higher decay-onset altitude $h_+$ could further suppress showers along steeper trajectories, concentrating the signal near the horizon and improving the angular agreement.
For context, the SM $\nu_\tau$ interpretation yields a $\gtrsim7\sigma$ ANITA-IV--IceCube tension for the diffuse scenario~\cite{Chattopadhyay:2026yoa}, while ANITA-IV quotes $\sim3.3\sigma$ for its four anomalous events against combined backgrounds~\cite{ANITA:2020gmv}. The mass flip-flop mechanism therefore offers a viable resolution of the ANITA-IV--IceCube tension.

Using the ANITA-IV flux normalization from Eq.~\eqref{eq:counts}, we also consider the first flight of PUEO~\cite{PUEO:2020bnn}.
We recalculate the effective area at PUEO's payload altitude of $35\,{\rm km}$~\cite{PUEO:2026owv} and apply a factor of $1.5$ to approximate its improved instrumental response compared to ANITA-IV.\footnote{We use the main-instrument air-shower projections of Ref.~\cite{Ku:2024Lv}, since the low-frequency trigger was removed before PUEO's first flight~\cite{Deaconu:2026PUEO}.} Accounting for the $12.9$ live days of its first flight~\cite{PUEO:2026owv}, we expect $2.7$ near-horizon events for BP1 and $3.0$ for BP2.
PUEO's first flight will thus provide an immediate test of the mass flip-flop mechanism.

\textbf{Summary and conclusions.}
The dark-sector mass flip-flop mechanism presented in this work provides a kinematic explanation for how a common incident flux can account for the four near-horizon ANITA-IV air-shower candidates while leaving no UHE signal at IceCube, in accord with observations. In the model, an ultralight scalar, in the presence of Earth matter, reverses the mass ordering of the two dark fermions, allowing an incident $N_1$ to decay into $N_2\,Z'$ deep underground. Both $N_1$ and $N_2$ decays in the vicinity of IceCube are blocked by construction. The $N_2$, stable in Earth matter, exits and decays in the atmosphere as $N_2\to N_1\, Z'$; the $Z'$ decays hadronically and initiates an air shower that produces a radio signal.

The nearly monoenergetic $N_1$ flux could originate from the two-body decay of EeV-scale dark matter. Fixing the flux for each benchmark to yield four ANITA-IV events, we predict about three near-horizon events in PUEO's first flight. 
Overall, we present a BSM scenario in which ANITA-IV has already observed a reversal of the dark-sector mass ordering induced by Earth itself.
PUEO, alongside other upcoming balloon-borne experiments, may bring us closer to establishing the BSM origin of these puzzling events.

\noindent
\textbf{Acknowledgments}.
C.A.A. are supported by the Faculty of Arts and Sciences of Harvard University, the National Science Foundation (NSF, CAREER Grant \#2239795), the John Templeton Foundation (Grant \#63651), the Research Corporation for Science Advancement, and the David \& Lucile Packard Foundation. The work of V.B. is supported by the United States Department of Energy Grant No.~DE-SC0016013.

\bibliographystyle{JHEP}
\bibliography{refs}

\clearpage
\newpage
\onecolumngrid
\setcounter{secnumdepth}{3}

\setcounter{equation}{0}
\setcounter{figure}{0}
\setcounter{table}{0}
\setcounter{page}{1}
\setcounter{section}{0}
\renewcommand{\thepage}{Supplemental Material -- S\arabic{page}}

\makeatletter
\renewcommand{\thesection}{S\arabic{section}}
\renewcommand{\theequation}{S\arabic{equation}}
\renewcommand{\thefigure}{S\arabic{figure}}
\renewcommand{\thetable}{S\arabic{table}}

\renewcommand{\theHfigure}{S\arabic{figure}}
\renewcommand{\theHtable}{S\arabic{table}}
\renewcommand{\theHequation}{S\arabic{equation}}
\makeatother

\begin{center}
    \textbf{\large Supplemental Material:\\ 
    \vspace{0.1cm}
    A Dark-Sector Mass Flip-Flop Mechanism for the ANITA-IV Events} \\ 
    \vspace{0.5cm}
    {Dibya~S.~Chattopadhyay, Carlos~A.~Arg\"uelles, Vedran~Brdar}
\end{center}

\section{Decay rates and energy transport}
\label{app:1}
The decay rates and the associated energies determine whether an incident $N_1$ produces a visible atmospheric shower through the $N_1 \to N_2 \to {\rm EAS}$ process. The rest-frame decay width for $N_i\to N_j\,Z'$ is~\cite{Djouadi:2001fa,Aguilar-Saavedra:2013qpa,Serao:2026qod}
\begin{align}
\Gamma_{ij}&\,=
\frac{|g_{12}|^2\sqrt{\lambda(M_i^2,M_j^2,M_{Z'}^2)}}
{16\pi M_i^3M_{Z'}^2}\,\Big[(M_i-M_j)^2-M_{Z'}^2\Big]
\Big[(M_i+M_j)^2+2M_{Z'}^2\Big],
\label{eq:width}
\end{align}
when $M_i>M_j+M_{Z'}$. Here, $\lambda(a,b,c)=a^2+b^2+c^2-2ab-2ac-2bc$.
At ultra-high energies, $E_i\gg M_i$, the laboratory decay length is $\ell_i(s,E_i)\simeq E_i/[M_i(s)\Gamma_{ij}(s)]$.
The survival probability of $N_i$ between positions $s_a$ and $s_b$ along its trajectory is
\begin{equation}
P_i(s_a,s_b;E_i)=
\exp\left[-\int_{s_a}^{s_b}
ds \,\ell^{-1}_i(s,E_i)\right],
\label{eq:survival}
\end{equation}
which accounts for the spatial variation of the matter-induced decay rate.

For the decay of an unpolarized $N_i$, in the ultrarelativistic limit, the energy fraction $z \equiv E_{Z'}/E_i$ is uniformly distributed in the $(z_-,\,z_+)$ interval, where~\cite{ParticleDataGroup:2024cfk,Agashe:2015ike}
\begin{equation}
z_\pm\simeq
\frac{M_i^2-M_j^2+M_{Z'}^2
\pm\sqrt{\lambda(M_i^2,M_j^2,M_{Z'}^2)}}
{2M_i^2}.
\label{eq:energy}
\end{equation}
The resulting dark fermion $N_j$ carries the remaining energy fraction of $1-z$. A monoenergetic incident $N_1$ flux therefore produces a range of exiting $N_2$ energies. The subsequent atmospheric decay $N_2\to N_1\,Z'$ introduces a further spread in the energy carried by the $Z'$ and hence in the energy of the EAS initiated by its hadronic decay. Our calculations incorporate the energy and position distributions of both decays, with each decay length evaluated for the corresponding particle energy.

For the $Z'$ decay, we use the hadronic widths of Ref.~\cite{Foguel:2022ppx}. The laboratory decay length is $\ell_{Z'}\simeq E_{Z'}/(M_{Z'}\Gamma_{Z'})$ and scales as $E_{Z'}g_B^{-2}$ for a fixed mass of $Z'$. We take $g_B=10^{-3}$, allowed by existing bounds in the relevant mass window around $M_{Z'}\sim\mathcal{O}(1)\,{\rm GeV}$~\cite{Tulin:2014tya,Dror:2017ehi,Foguel:2022ppx}. At $E_{Z'}=1\,{\rm EeV}$, the decay length remains below $0.1\,{\rm km}$ for $M_{Z'}\gtrsim0.7\,{\rm GeV}$~\cite{Foguel:2022ppx,Foguel:2024lca}. Such length scales are much smaller than the atmospheric ones relevant for shower development. We therefore treat the hadronic decay, with branching fraction $B_h\simeq1$~\cite{Foguel:2022ppx}, as effectively instantaneous, placing the shower origin at the $N_2$ decay position and taking the EAS energy to be $E_{\rm sh}\simeq E_{Z'}$.

\section{Effective area}
\label{app:2}
The effective area of a detector determines its instantaneous sensitivity to an incident particle of a given energy and arrival direction~\cite{ANITA:2021xxh,Chattopadhyay:2026yoa}.
For ANITA-IV, the effective area in our flip-flop setup (denoted in what follows with `BSM') combines the probability that an incident $N_1$ produces an emerging $N_2$, the probability that the $N_2$ decays in the atmosphere, and the detector response to the resulting shower.
We adopt the latter from Ref.~\cite{Chattopadhyay:2026yoa}, matching the visible shower energies and correcting for the different branching ratios and decay locations. This prescription takes the form
\begin{align}
A_{\rm eff}^{\rm BSM}(E_1,\theta)
=&\,A_{\rm geo} \otimes P_{\rm exit}^{\rm BSM} \otimes P_{\rm decay}^{\rm BSM}\otimes P_{\rm trig}^{\rm BSM} \,.
\label{eq:area}
\end{align}
Here $A_{\rm geo}$ is the geometric area viewed by the payload and $\theta$ is the elevation angle. The exit probability $P_{\rm exit}^{\rm BSM}$ includes $N_1\to N_2\,Z'$ conversion in Earth;  $P_{\rm decay}^{\rm BSM}$
gives the probability of $N_2\to N_1\,Z'$ along the atmospheric path considered. Further, we define
\begin{equation}
    P_{\rm trig}^{\rm BSM}(E_1,\theta) = \eta_{\rm EAS} \,\eta_{\rm BR} \,P_{\rm trig}^{\tau}(E_{\nu,\rm cal},\theta) \, ,
    \label{eq:ptrig}
\end{equation}
where $\eta_{\rm EAS}\equiv F_{\rm dev}^{\rm BSM}/F_{\rm dev}^{\tau}$ corrects for the different fractions of atmospheric decays that occur early enough for the shower to develop, while $\eta_{\rm BR}\equiv B_h/B_{\tau,\rm sh}$ accounts for the different branching fractions into shower-producing final states. Namely, the hadronic branching fraction for a $U(1)_B$ $Z'$ is $B_h=1$~\cite{Foguel:2022ppx,Foguel:2024lca}, while for a $\tau$ lepton, considering both hadronic and electronic tau decays~\cite{ParticleDataGroup:2024cfk}, we obtain
$B_{\tau,\rm sh}=0.826$. For each angular direction, $E_{\nu,\rm cal}$ is the incident $\nu_\tau$ energy that matches the SM mean visible shower energy to the BSM value at $E_1$, with both means evaluated after the air shower development cut.
We consider a $1.5^\circ$ viewing cone~\cite{Chattopadhyay:2026yoa}, with each path assigned its own Earth chord, exit point, and atmospheric path.

To calculate the EAS development fraction $F_{\rm dev}$, for a decay at distance $s$ from the surface, we calculate the remaining air column, $X(s)=\int_s^L\rho[h(u)]\,du$, using a layered atmospheric profile, where $L$ is the path length to the trajectory's closest approach to the payload.
As a first approximation to the shower-development requirement, we require $X(s)\geq700\,{\rm g\,cm^{-2}}$~\cite{MammenAbraham:2022xoc} and define the ratio
\begin{equation}
F_{\rm dev} = \frac{\int_0^{S_\star} p_{\rm dec}(s)\,ds}{\int_0^{L} p_{\rm dec}(s)\,ds}\,,
\label{eq:fdev}
\end{equation}
where $S_\star$ is the largest decay distance satisfying $X(S_\star)\geq700\,{\rm g\,cm^{-2}}$ and $p_{\rm dec}(s)$ is the decay probability density averaged over the energy spectrum of the exiting particle.\footnote{We use this sharp slant-depth cut as a first approximation for estimating the effective area~\cite{PierreAuger:2014sui,Cummings:2021bhg,MammenAbraham:2022xoc}; a more realistic treatment would require modeling of the radio signal from showers at different stages of development and the resulting detector response~\cite{Romero-Wolf:2018zxt,ANITA:2021xxh}.}
Our semi-analytic, phenomenological prescription reproduces the reference SM effective areas reasonably well~\cite{ANITA:2021xxh,Chattopadhyay:2026yoa} and allows us to extend the analysis to the $N_1 \to N_2 \to {\rm EAS}$ scenario, accounting for its different decay pathways and visible shower energies.

\end{document}